\documentclass{aa}  
\usepackage{orcidlink}
\usepackage{natbib}
\bibpunct{(}{)}{;}{a}{}{,} 
\usepackage{graphicx}
\usepackage{txfonts}
\usepackage{color}
\usepackage{threeparttable}
\usepackage{booktabs}
\usepackage{caption}
\newcommand{\nicer}{\emph{NICER}}

\begin{document}

\title{Stable X-ray reverberation lags in the black hole X-ray binary Swift J1727.8--1613}

\author{
Wei Yu\inst{1}\thanks{wei.yu@mnf.uni-tuebingen.de}
\orcidlink{0000-0002-3229-2453},
Sinan Allak\inst{1},
Zi-Xu Yang\inst{2}\orcidlink{0000-0003-1718-8487},
Xiao Fan\inst{3}\orcidlink{0000-0001-7350-8380},
Andrea Santangelo\inst{1}\orcidlink{0000-0003-4187-9560},
Tian-hao Xie\inst{4,5,6}
}

\institute{
Institut f\"ur Astronomie und Astrophysik, Kepler Center for Astro and Particle Physics, Eberhard Karls Universit\"at, Sand 1, 72076 T\"ubingen, Germany
\and School of Physics and Optoelectronic Engineering, Shandong University of Technology, Zibo 255000, China
\and Department of Astronomy, School of Physics and Technology, Wuhan University, Wuhan 430072, People's Republic of China
\and Institute of Large-scale Scientific Facility and Centre for Zero Magnetic Field Science, Beihang University, Beijing, 100191, China
\and School of Instrumentation Science and Opto-electronics Engineering, Beihang University, Beijing, 100191, China
\and National Institute of Extremely-Weak Magnetic Field Infrastructure, Hangzhou, 310051, China
}

\abstract
{}
{We investigate the evolution of X-ray reverberation lags in the black hole X-ray binary Swift J1727.8--1613 during its 2023 outburst, with the aim of probing the inner accretion flow geometry across spectral states.}
{We analyzed \nicer{} observations covering the low-hard state (LHS) and hard-intermediate state (HIMS). The time lags were computed using Fourier-based techniques, and we constructed lag--frequency and lag--energy spectra. To obtain a robust estimate of the soft lag amplitude, we focused on a frequency range in which the reverberation signal dominates and remains stable, thereby minimizing contamination from hard lags and phase-wrapping effects.}
{The soft lag amplitude increases rapidly from the LHS to the early HIMS and then stays near $\sim$10 ms throughout the HIMS. In the frequency range in which reverberation lags prevail, the lag shows little dependence on Fourier frequency. On the other hand, the observed low-frequency lags change from hard-lag dominated to soft-lag dominated, with amplitudes comparable to those measured at higher frequencies.}
{These results suggest that the reverberation lag varies little during the HIMS, consistent with a relatively stable inner accretion geometry during this state. The apparent evolution of the lag amplitude from the LHS to the HIMS can be largely explained by the diminishing effect of hard lags and does not necessarily require significant changes in the intrinsic light-travel timescale. Swift J1727.8--1613 therefore provides a case in which the reverberation signal can be studied with reduced contamination over a broad frequency range, offering new insight into the evolution of the accretion geometry in black hole X-ray binaries.}


\titlerunning{X-ray reverberation lags in Swift J1727.8--1613}
\authorrunning{Wei Yu et al.}
\maketitle

\section{Introduction}

Black hole low-mass X-ray binaries (BH-LMXBs) are predominantly transient systems in which a black hole accretes matter from a companion star via an accretion disk \citep{shakura1973black}. These systems, commonly referred to as black hole transients (BHTs), spend most of their lifetime in quiescence and occasionally undergo outbursts lasting from weeks to months. Such outbursts are generally attributed to thermal-viscous instabilities in the accretion disk \citep{cannizzo1995accretion, lasota2001disc}. During these episodes, the source luminosity can increase dramatically, in some cases approaching the Eddington limit, and this is accompanied by pronounced changes in the spectral and timing properties. These variations allow us to classify distinct accretion states \citep{belloni2009states}.

The spectral evolution of BHTs during outbursts follows a characteristic hysteresis pattern in the hardness-intensity diagram (HID). As the source emerges from quiescence, it typically enters the low-hard state (LHS), where the X-ray emission is dominated by a hard nonthermal component produced via inverse-Compton scattering of soft-disk photons by hot electrons in a corona. The spectrum in this state is well described by a power law with a high-energy cutoff \citep{zdziarski2004radiative, remillard2006x, done2007modelling}. The corresponding power density spectrum (PDS) typically shows strong band-limited noise and prominent low-frequency quasi-periodic oscillations (LFQPOs). As the luminosity increases, the source evolves through intermediate states, that is, the hard-intermediate state (HIMS) and the soft-intermediate state (SIMS), before it reaches the high-soft state (HSS), where the emission is dominated by a geometrically thin optically thick accretion disk and can be modeled with a multitemperature blackbody \citep{remillard2006x, you2016testing}. These state transitions are closely associated with changes in the timing properties, particularly, with the evolution of LFQPO types: type-C QPOs are typically observed in the LHS and HIMS, while type-B and type-A QPOs are characteristic of the SIMS and HSS \citep{casella2005abc, ingram2019review}.

X-ray variability studies further revealed complex time delays between different energy bands \citep{uttley2025large}. At low Fourier frequencies (typically below $\sim$1 Hz), harder X-ray photons are observed to lag softer ones, producing the so-called hard lags. These lags are widely interpreted as a consequence of inward-propagating fluctuations in the mass-accretion rate in the disk and corona that modulate emission at progressively higher energies as they reach the inner regions \citep{kotov2001x, arevalo2006investigating}. Additional contributions might arise from Comptonization processes, where higher-energy photons undergo more scattering events and thus experience longer delays \citep{kylafis2008jet}. 

At higher Fourier frequencies, however, a different type of delay can emerge. A fraction of the Comptonized photons irradiates the optically thick accretion disk and is reprocessed, producing a delayed response in the soft X-ray emission. This leads to soft X-ray lags, where the soft band follows variations in the hard band on short timescales. This is commonly referred to as X-ray thermal reverberation \citep[e.g.,][]{uttley2011causal,de2015tracing}. As the reverberation signal is set by the light-travel time between the X-ray source and the disk, such lags provide a direct probe of the geometry of the inner accretion flow. Recently, \citet{zhan2025modeling} implemented the propagating-fluctuation and reverberation scenarios within an accretion disk--corona system using Monte Carlo radiative transfer simulations. This framework enables a quantitative study of rapid X-ray variability and naturally reproduces the low-frequency hard lags and high-frequency soft lags observed in lag--frequency spectra, successfully explaining the reverberation properties of MAXI J1820+070 \citep{kara2019corona}.

In recent years, X-ray reverberation lags have been widely detected in black hole X-ray binaries (BHXBs), and their evolution has been suggested to correlate with accretion states \citep{uttley2011causal,de2015tracing, de2017evolution, marco2016reverberation, wang2020relativistic, wang2022nicer, you2026reverberation}. Since reverberation lags are primarily determined by light-crossing times, they provide a model-independent probe of the inner accretion geometry. Previous studies showed that the amplitude and characteristic frequency of reverberation lags can vary during outbursts, and these changes are often interpreted in terms of variations in the inner accretion flow, such as a changing disk-truncation radius or coronal structure \citep[e.g.,][]{kara2019corona,de2021inner,wang2022nicer,yu2023spectral}.

However, the measurement and interpretation of reverberation lags remain challenging in practice. The observed lag signals are frequently affected by the coexistence of multiple variability components, particularly, contamination from low-frequency hard lags and dilution effects due to mixed spectral contributions. The relative importance of these components can vary across spectral states, making it difficult to isolate the intrinsic reverberation signal, especially during intermediate states in which different variability processes overlap \citep{you2026reverberation}. In addition, reverberation lags in BHXBs typically occur on millisecond timescales, requiring observations with sufficiently high count rates and excellent timing capabilities to be reliably detected. This makes bright outbursts particularly valuable for studying the intrinsic properties of reverberation signals.

The X-ray transient Swift J1727.8--1613 was discovered by the Burst Alert Telescope (BAT) aboard the \textit{Neil Gehrels Swift Observatory} (\emph{Swift}) on August 24, 2023 \citep{page2023grb}. A rapid increase in flux confirmed it as a new Galactic X-ray transient \citep{negoro2023maxi, nakajima2023maxi}. Observations in multiple wavelengths, including the optical \citep{castro2023optical}, X-ray \citep{o2023nicer}, and radio \citep{miller2023vla}, suggested that Swift J1727.8--1613 is a low-mass black hole candidate. The distance of this source was estimated to be $D = 3.4\pm0.3$ kpc by \citet{sanchez2025dynamical} and $5.5^{+1.4}_{-1.1}$ kpc by \citet{burridge2025distance}. An extremely high spin value of 0.98 was obtained from reflection modeling \citep{liu2026broad}. The temporal and spectral analyses by \citet{yu2024timing} and \citet{chatterjee2024insight} suggest a high-inclination accretion disk for this source. The X-ray polarization properties of the source were monitored by the Imaging X-ray Polarimetry Explorer (IXPE) \citep{veledina2023discovery,svoboda2024dramatic,ingram2024tracking}, with the polarization measurements suggesting an inclination angle of \(i \sim 30^\circ\!-\!60^\circ\) and a distance of \(\sim 1.5\)  kpc. Timing studies revealed strong low-frequency QPOs throughout the outburst \citep{yang2023fast, yu2024timing, liao2025tracking, cao2025spectral, xu2025temporal}. The phase-dependent behavior of the X-ray polarization with respect to the QPO has also been investigated \citep{zhao2024first}.
During the outburst, the source reached a peak brightness of $\sim$8 Crab, making it an exceptional target for reverberation studies \citep{palmer2023swift}. \citet{ingram2024tracking} reported that the evolution of the soft-lag amplitude in this source does not follow the common trend identified in previous population studies \citep{wang2022nicer}. 

We perform a systematic analysis of the lag spectra of Swift J1727.8--1613 using The Neutron star Interior Composition Explorer (\nicer{}) observations obtained during its 2023 outburst, with the aim of investigating the evolution of the inner accretion flow geometry across spectral states. The paper is structured as follows. Sect.~\ref{sec2} describes the observations and data reduction. The results are presented in Sect.~\ref{sec3}, followed by the discussion and conclusions in Sects.~\ref{sec4} and \ref{sec5}.

\section{Observations and data reduction}\label{sec2}

\subsection{NICER}

\nicer{} is an X-ray timing instrument on board the International Space Station (ISS), providing high throughput and excellent time resolution in the 0.2--12 keV energy range. These capabilities make it particularly well suited for timing studies of accreting compact objects.

We analyzed \nicer{} observations of Swift J1727.8--1613 obtained during its 2023 outburst, covering the LHS and HIMS \citep{stiele2024nicer}. The observations spanned the period from August 25 to September 16, 2023, and were retrieved from the HEASARC archive. To ensure reliable time-lag measurements, we applied an initial screening and retained only observations with exposure times longer than 100 s.

The \nicer{} event data can be affected by numerous short good time intervals (GTIs), a phenomenon often referred to as GTI shredding, which is typically associated with high count rates or overly strict screening criteria. Given the high flux of Swift J1727.8--1613, this effect is particularly pronounced and can hinder a reliable timing analysis. To mitigate this issue, we reprocessed the data using the \texttt{nicerl2} pipeline from HEASoft (v6.32.1), which incorporates improved filtering procedures designed to reduce GTI fragmentation.

\subsection{Time-lag measurements}

We measured time lags between different energy bands using standard Fourier-based techniques \citep{nowak1999rossi, vaughan2003characterizing, uttley2014x}. The light curves were extracted in selected energy bands with a time resolution of 1 ms. The discrete Fourier transform (DFT) was computed for each light curve, and the cross-spectrum was obtained as the product of the Fourier transform of one band and the complex conjugate of another. 

The cross-spectra were averaged over logarithmically spaced frequency bins. The phase lag at each frequency was then converted into a time lag by dividing by $2\pi \nu$, where $\nu$ is the Fourier frequency. Throughout this work, a positive lag indicates that variations in the hard X-ray band lag behind those in the soft band. Since no prominent Fe K reverberation feature was detected in the lag--energy spectra, we followed \citet{kara2019corona} and adopted the 0.5--1 keV and 1--10 keV bands as the soft and hard bands, respectively, for the calculation of lag--frequency spectra to maximize the signal-to-noise ratio. All timing analyses were performed using the software package \texttt{Stingray} \citep{matteo_bachetti_2026_19002995}, which provides a comprehensive framework for an X-ray timing analysis.

\subsection{Measurement of the soft-lag amplitude}

To quantify the amplitude of soft lags, a commonly adopted approach is to compute an average lag over a selected frequency range in which the soft lags dominate. In this approach, the frequency interval is typically defined between a lower bound $\nu_{\rm min}$ and an upper bound $\nu_{\rm max}$, excluding frequencies affected by quasi-periodic oscillations. The lower bound is often chosen as the lowest frequency at which the lag becomes negative, and the upper bound is set to the critical frequency for phase wrapping.

However, the determination of $\nu_{\rm min}$ and $\nu_{\rm max}$ can be affected by the coexistence of multiple lag components. In particular, $\nu_{\rm min}$ might be affected by low-frequency hard lags, such that stronger hard-lag contributions can shift $\nu_{\rm min}$ to higher frequencies, potentially biasing the measured soft-lag amplitude. Similarly, although $\nu_{\rm max}$ is often associated with the characteristic frequency of phase wrapping, its effect on the measured lag amplitude depends on the underlying impulse response. In the case of an idealized delta-function response, phase wrapping does not affect the lag amplitude below $\nu_{\rm max}$. In more realistic scenarios involving extended responses, however, the effect of phase wrapping can already become significant at frequencies below $\nu_{\rm max}$, leading to an underestimation of the soft-lag amplitude \citep{uttley2014x}.

For Swift J1727.8--1613, we adopted a more conservative fixed frequency interval of 1--10 Hz to minimize low-frequency hard-lag and phase-wrapping effects. Within this range, the reverberation component is generally dominant and depends only weakly on the frequency in most observations. The same interval for all epochs also ensured a homogeneous comparison of the lag evolution throughout the outburst. We note that the qualitative evolution of the soft-lag amplitude is not sensitive to modest adjustments of the adopted frequency interval within this range.

\begin{table*}[htbp]
\centering
\captionsetup{justification=centering}
\caption{Representative \nicer{} observations of Swift J1727.8--1613.}
\begin{tabular}{cccccc}
\toprule
Epoch & ObsID & Time (UTC) & MJD & Exposure (s) & State\\
\midrule
Epoch 1 & 6203980101 & 2023-08-25 00:23:40.00 & 60181 & 487 & LHS \\
Epoch 2 & 6203980104 & 2023-08-28 01:26:40.00 & 60184 & 3125 & LHS \\
Epoch 3 & 6203980106 & 2023-08-30 00:00:20.00 & 60186 & 10585 & HIMS \\
Epoch 4 & 6203980109 & 2023-09-02 00:40:40.00 & 60189 & 12580 & HIMS \\
Epoch 5 & 6750010301 & 2023-09-06 00:44:40.00 & 60193 & 7349 & HIMS \\
Epoch 6 & 6703010107 & 2023-09-15 06:27:26.00 & 60202 & 4529 & HIMS \\
\bottomrule
\end{tabular}
\label{table1}
\end{table*}

\section{Results}\label{sec3}

\begin{figure*}[htbp]
    \includegraphics[width=\columnwidth]{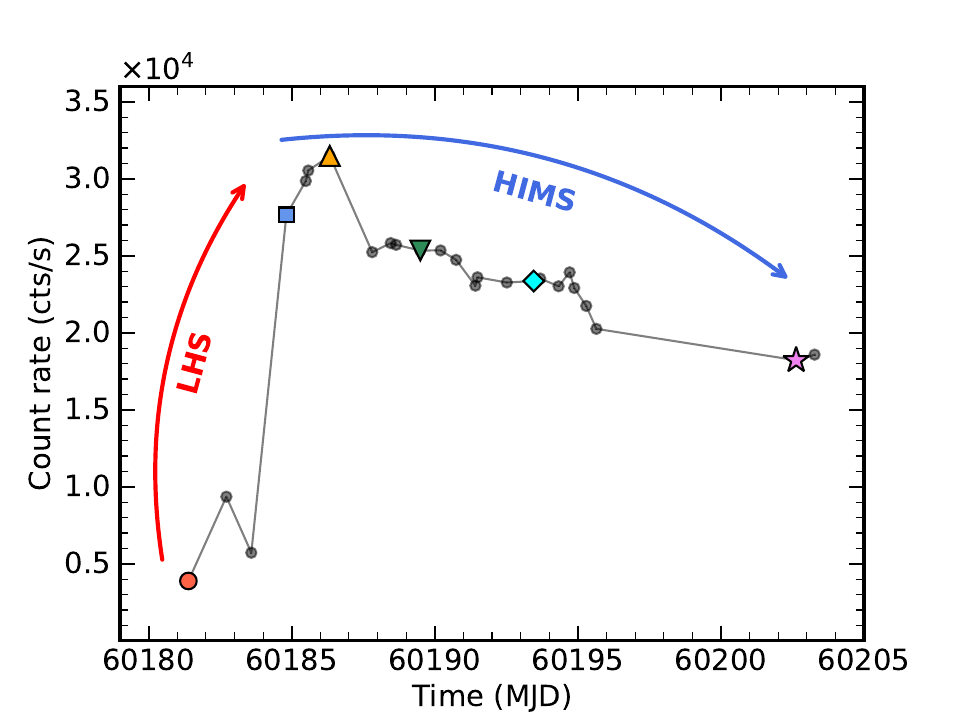}
    \includegraphics[width=\columnwidth]{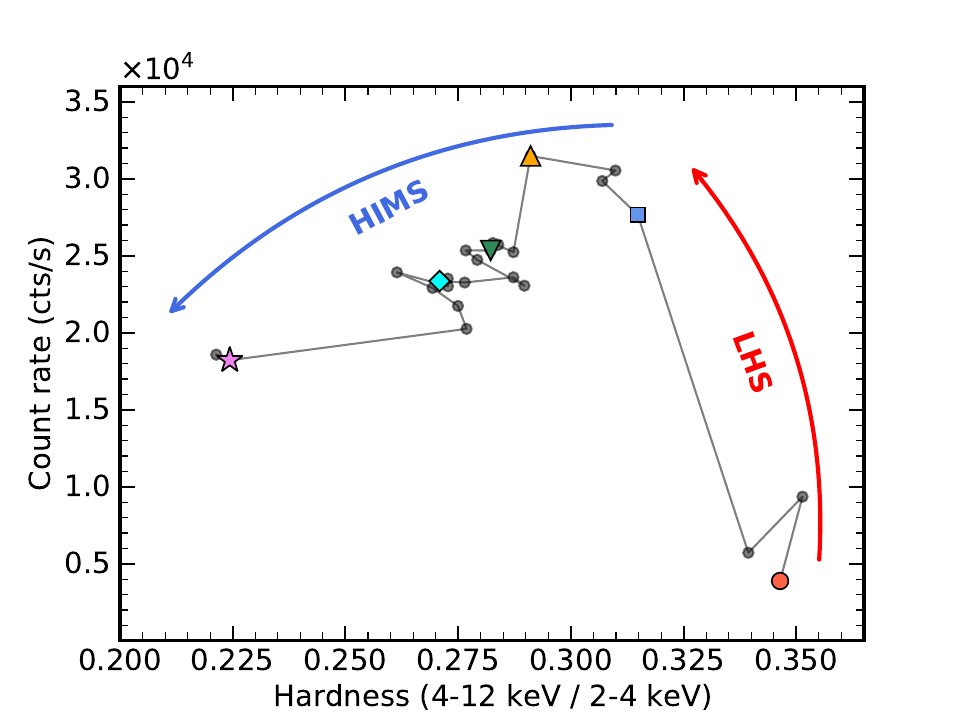}
    \caption{Long-term evolution of Swift J1727.8--1613 during its 2023 outburst. \textit{Left:} \nicer{} light curve showing the count rate as a function of time (MJD). The six colored markers indicate the representative epochs selected for the timing analysis. \textit{Right:} Hardness--intensity diagram, where the hardness is defined as the count rate ratio of the 4--12 keV and 2--4 keV bands. The source follows the typical evolution from the low hard state (LHS) toward the hard intermediate state (HIMS), as indicated by the track in the HID. The selected epochs span this transition and were used for our detailed timing analysis.}
    \label{fig1}
\end{figure*}

\begin{figure}
        \includegraphics[width=\columnwidth]{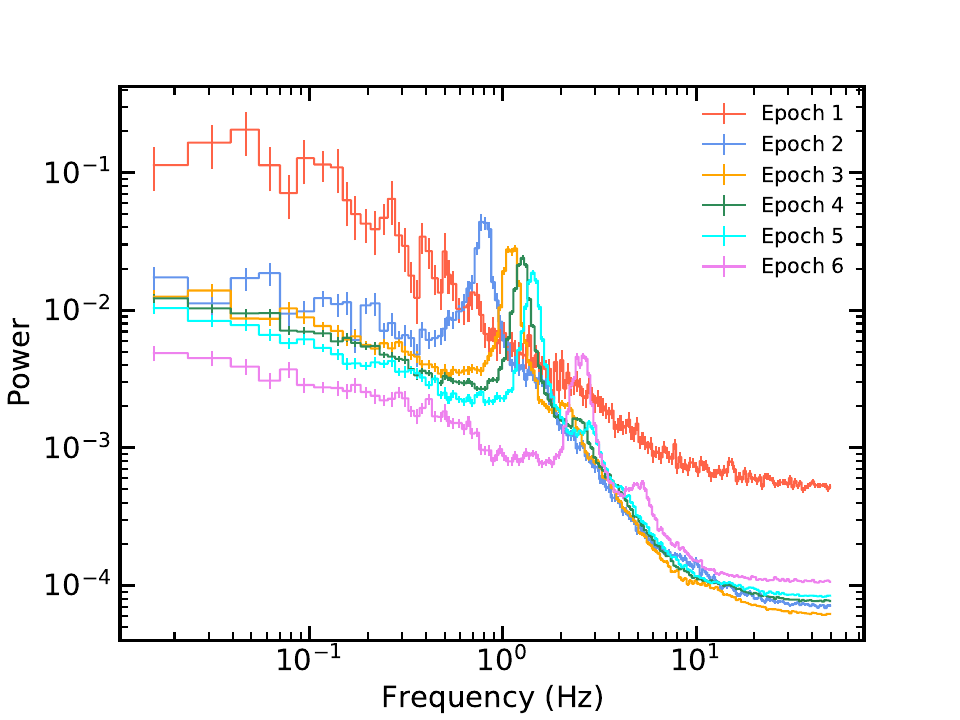}
    \caption{Power density spectra of the six representative epochs shown in Fig.~\ref{fig1}. Strong band-limited noise is observed in all epochs, with a clear evolution in amplitude and characteristic frequency. Except for epoch 1, prominent type-C QPOs are detected, with centroid frequencies shifting toward higher values from epoch 2 to epoch 6.}
    \label{fig2}
\end{figure}

\subsection{Fundamental diagrams}

Figure~\ref{fig1} shows the long-term \nicer{} light curve and HID of Swift J1727.8--1613 during its 2023 outburst. The light curve initially rose rapidly, followed by a plateau phase during which the count rate remained relatively stable for about ten days. After a short observational gap, the source was detected at a lower count rate during the decay phase.

The HID displays a partial q-shaped track commonly observed in BHXBs, with an overall softening of the source as the outburst progresses. The observations we analyzed cover the initial rise and the upper branch of the HID, corresponding to the transition from the typical LHS to the HIMS \citep{belloni2009states, stiele2024nicer}.
To facilitate the subsequent timing analysis, we selected six representative epochs of the observations, which are marked in Fig.~\ref{fig1}. The details of these observations are listed in Table~\ref{table1}.

For each epoch, we computed the PDS to characterize the variability properties. Except for the first epoch, strong type-C QPOs are detected, with centroid frequencies shifting toward higher values from epoch 2 to epoch 6. This is consistent with the typical behavior observed during the transition from the LHS to the HIMS.

\subsection{Lag--frequency spectra}

To explore the evolution of time lags, we computed lag--frequency spectra for all epochs. The results are shown in Fig.~\ref{fig3}. In general, the lag spectra exhibit three distinct regimes commonly observed in BHXBs. At low Fourier frequencies, the lags are dominated by hard lags, with the amplitude decreasing as a function of frequency. At intermediate frequencies, soft lags become prominent, and at higher frequencies, the lag spectra oscillate due to phase wrapping.

In the LHS (epochs 1 and 2), the lag spectrum shows clear hard lags below $\sim$1 Hz, consistent with the behavior observed in most BHXBs. Such hard lags are commonly interpreted as arising from the inward propagation of fluctuations in the mass-accretion rate. At higher frequencies ($\sim$1--10 Hz), the lag spectrum becomes dominated by soft lags, which are typically attributed to reverberation delays. Unlike the commonly observed reverberation-lag evolution in BHXBs, as the source evolves toward the HIMS, the low-frequency hard lags decrease rapidly \citep{de2015tracing, kara2019corona}. At the same time, the soft lags increase from $\sim$2 ms in the early LHS to $\sim$10 ms and remain approximately constant in subsequent epochs. Notably, this stability is observed not only across different observations, but also over a broad frequency range in which soft lags dominate.

To quantify the evolution of the soft lags, we measured the average lag amplitude in the 1--10 Hz frequency range, where the soft lag dominates and the lag--frequency spectrum remains approximately flat. To minimize the effect of QPO-related lags, we excluded the frequency range corresponding to the full width at half maximum (FWHM) of the QPO. The results are shown in Fig.~\ref{fig4}, where we plot the absolute amplitude of the negative soft lag for clarity. The soft-lag amplitude increases rapidly during the transition from the LHS to the HIMS and then remains nearly constant throughout the HIMS. This behavior differs from the commonly observed trend in other black hole X-ray binaries, where reverberation lags are reported to decrease gradually during the HIMS \citep[e.g.,][]{kara2019corona,de2021inner,wang2022nicer,yu2023spectral}.

We also measured the average lag in the low-frequency range of 0.1--1 Hz, where hard lags typically dominate. The results are shown in Fig.~\ref{fig5}. The hard lag decreases rapidly from the LHS and approaches zero, showing an evolution opposite to that of the high-frequency soft lags. As the source evolves into the HIMS, the lag becomes negative, marking a transition from hard to soft lags. Notably, the soft-lag amplitude is comparable to that measured in the 1--10 Hz range (Fig.~\ref{fig4}), with typical values of $\sim$10 ms, indicating that reverberation lags dominate over a broad frequency range during the HIMS.

\subsection{Lag--energy spectra}

To investigate the energy dependence of the time lags, we computed lag--energy spectra in the two representative frequency ranges: a low-frequency range (0.1--1 Hz), where hard lags are expected to dominate, and a high-frequency range (1--10 Hz), where soft (reverberation) lags are prominent. The lag--energy spectra were obtained by measuring the lag between a series of narrow energy bands and a broad reference band. We adopted the 1--10 keV band as the reference band, excluding the energy channel of interest to avoid correlated noise. The resulting lag--energy spectra are shown in Figs.~\ref{fig6} and \ref{fig7} for the high- and low-frequency ranges, respectively.

\begin{figure}
        \includegraphics[width=\columnwidth]{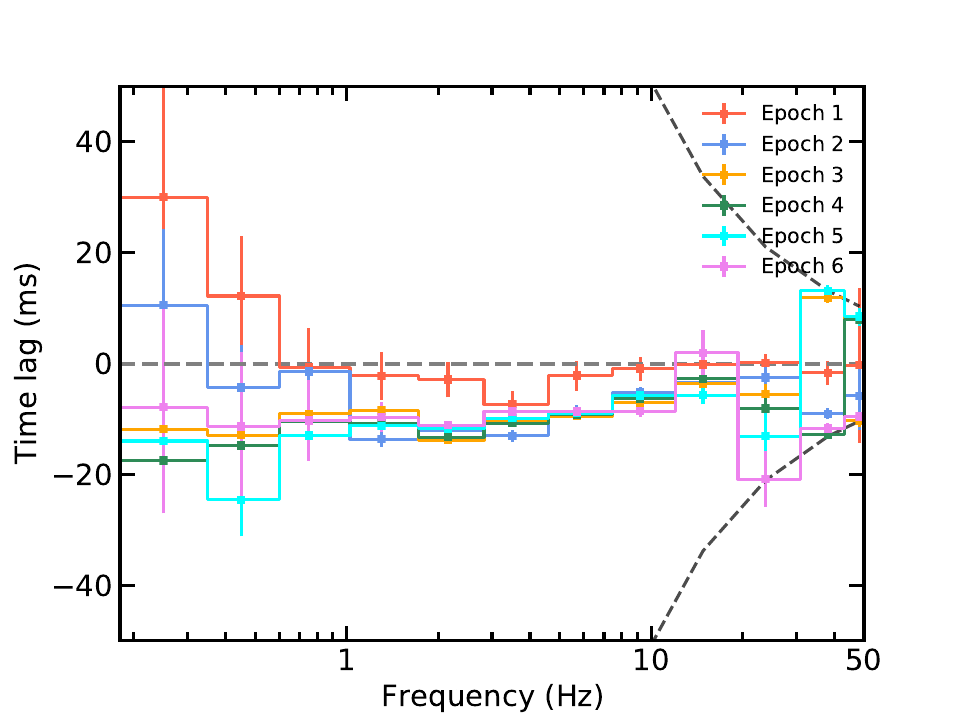}
    \caption{Lag--frequency spectra of the six representative epochs, computed between the 0.5--1 keV (soft) and 1--10 keV (hard) bands. Positive lags indicate that the hard photons lag the soft ones. The spectra show hard lags at low frequencies ($\lesssim$1 Hz), soft lags at intermediate frequencies ($\sim$1--10 Hz), and phase-wrapping features at higher frequencies. The soft-lag amplitude increases from epoch 1 to epoch 2 and remains approximately constant thereafter. The dashed gray  lines indicate the phase-wrapping limits.}

    \label{fig3}
\end{figure}

\begin{figure}
        \includegraphics[width=\columnwidth]{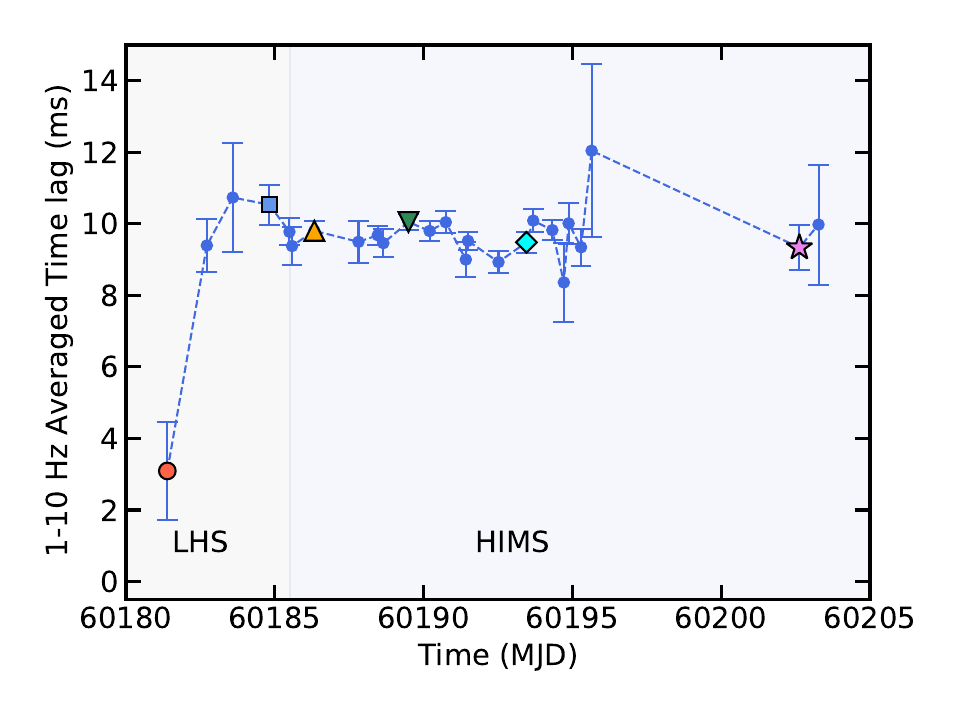}
    \caption{Evolution of the average soft-lag amplitude in the 1--10 Hz frequency range as a function of time. For clarity, the sign of the lag has been inverted so that higher values correspond to larger soft-lag amplitudes. The soft lag increases rapidly from the LHS to the early HIMS and remains approximately constant thereafter. The colored markers indicate the six representative epochs shown in Fig.~\ref{fig1}.}

    \label{fig4}
\end{figure}

\begin{figure}
        \includegraphics[width=\columnwidth]{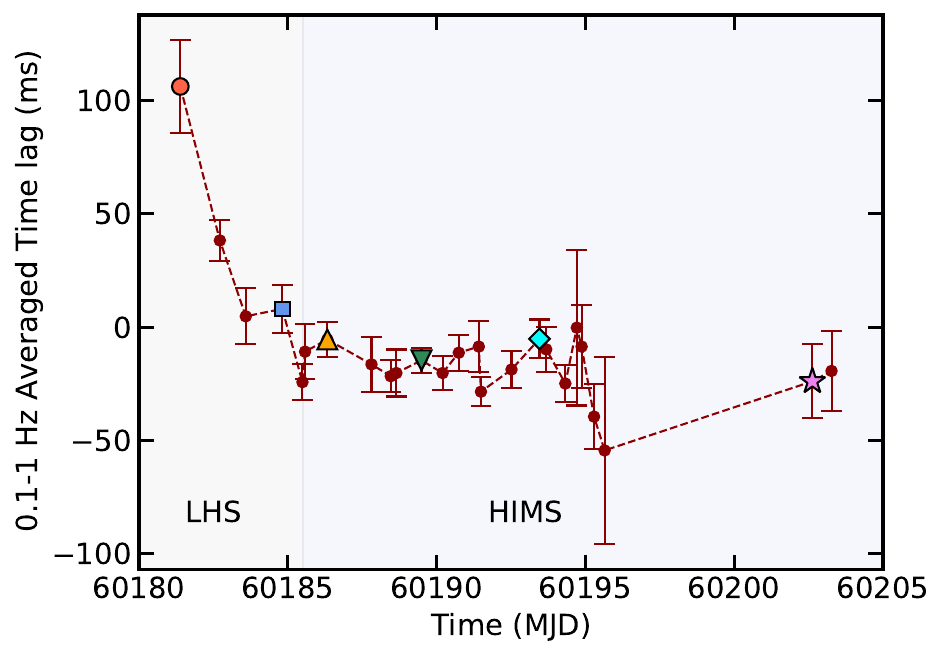}
    \caption{Evolution of the average lag in the 0.1--1 Hz frequency range as a function of time. Unlike Fig.~\ref{fig4}, the signed lag is shown here following the convention that positive values indicate hard lags. The lag changes from positive values (hard lags) in the LHS to negative values (soft lags) in the HIMS. The colored markers indicate the six representative epochs shown in Fig.~\ref{fig1}.}
    \label{fig5}
\end{figure}

\begin{figure*}
            \includegraphics[width=6.5cm]{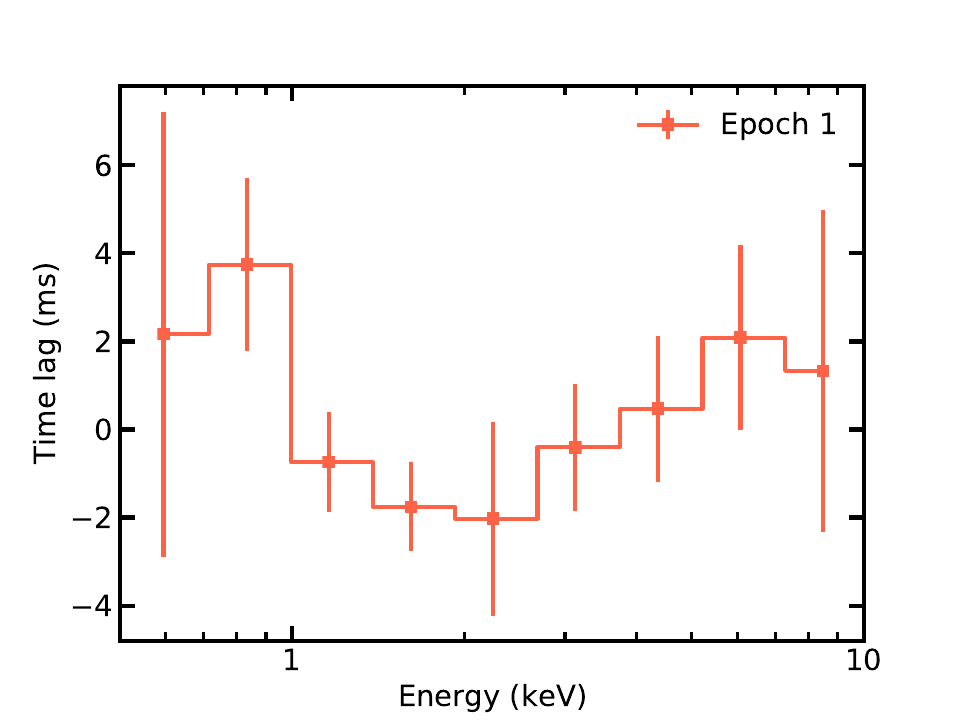}
        \includegraphics[width=6.5cm]{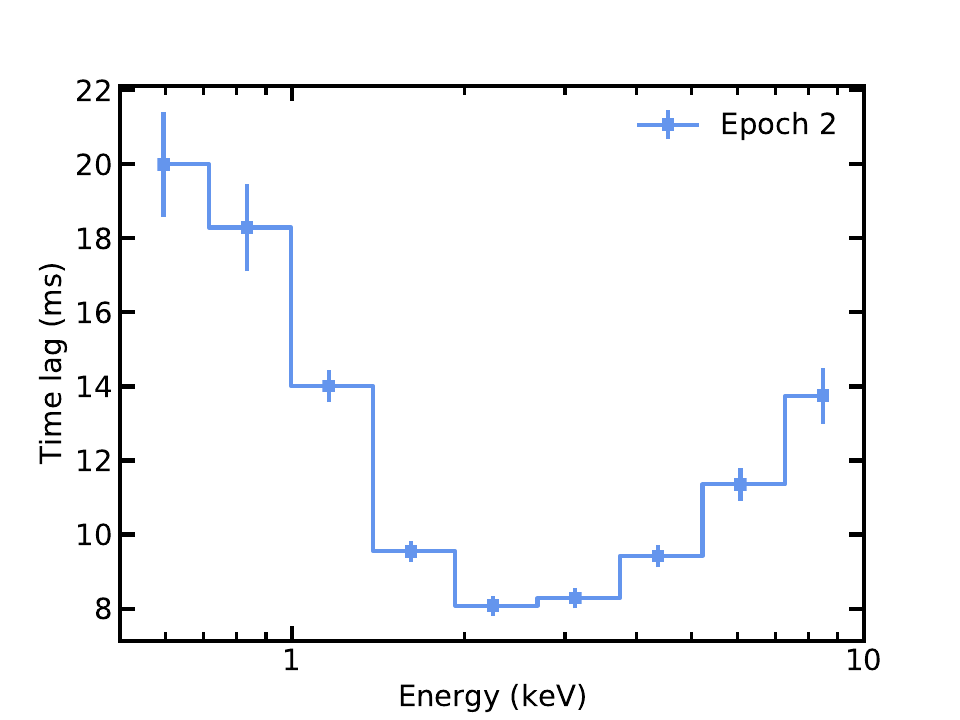}
        \includegraphics[width=6.5cm]{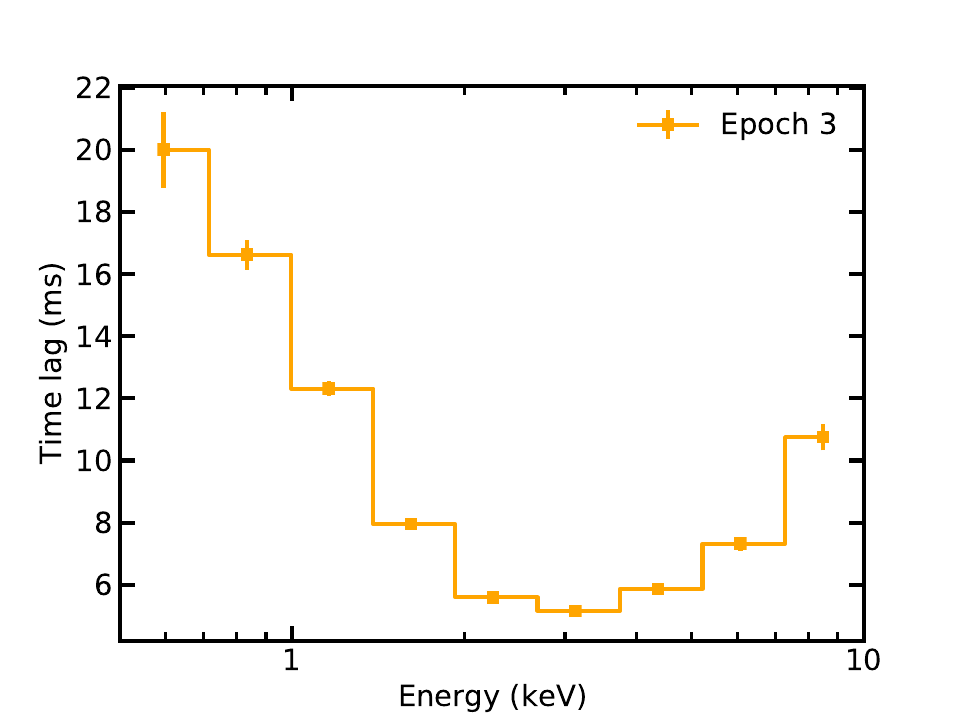}
        \includegraphics[width=6.5cm]{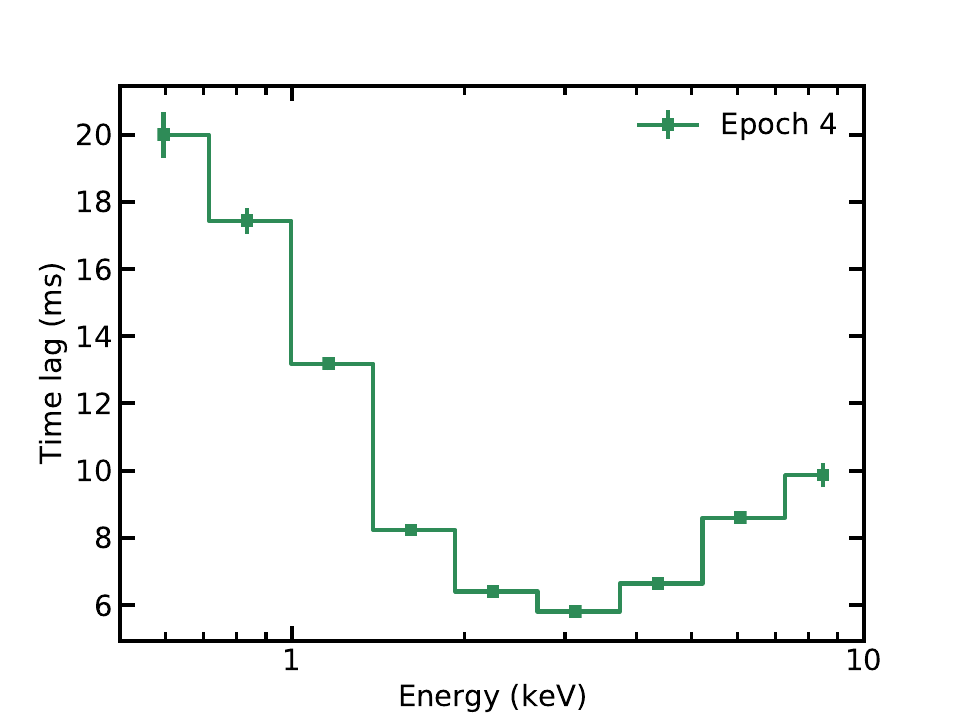}
        \includegraphics[width=6.5cm]{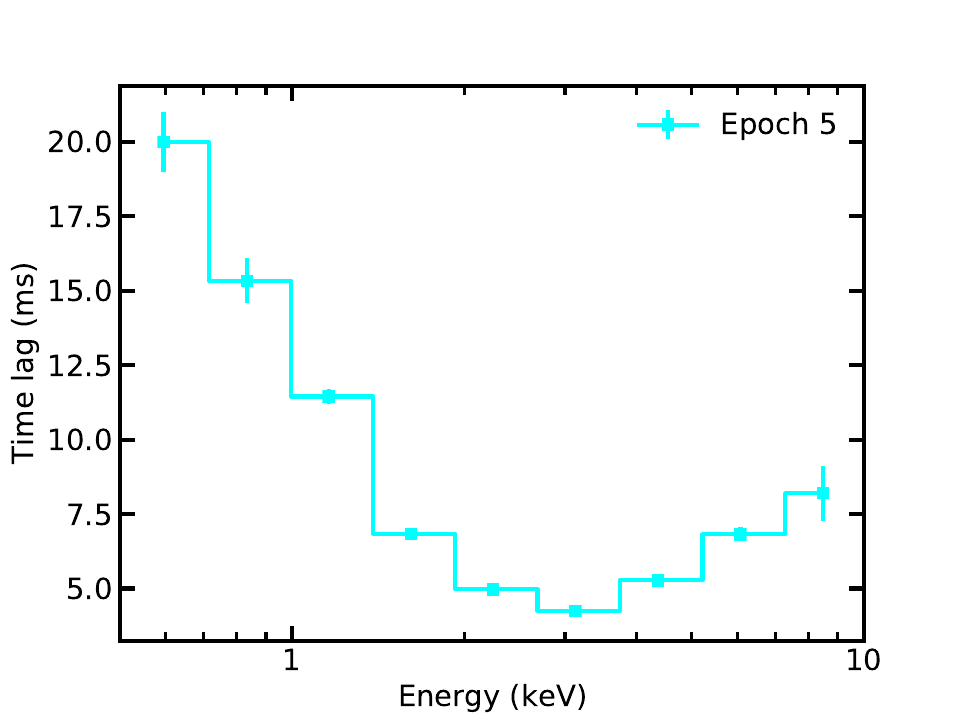}
        \includegraphics[width=6.5cm]{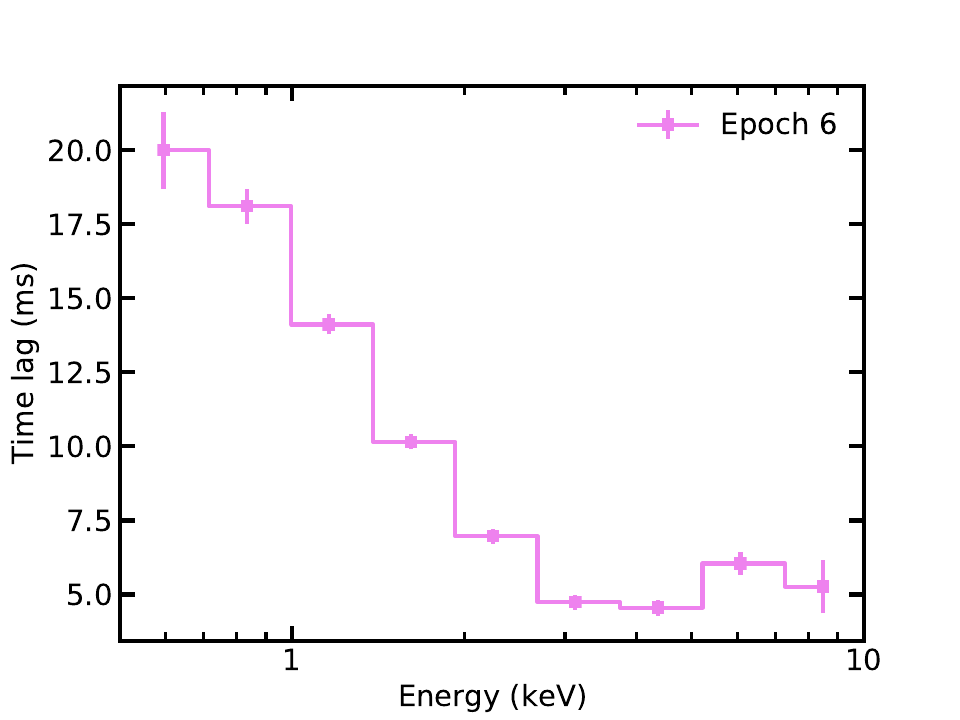}
    \caption{Lag--energy spectra in the 1--10 Hz frequency range for the six representative epochs. The lags are computed with respect to a 1--10 keV reference band, excluding the energy channel of interest. All epochs display a similar valley-like structure, with the overall shape remaining stable while the amplitude increases from epoch 1 to epoch 2 and remains approximately constant thereafter.}

    \label{fig6}
\end{figure*}

\begin{figure*}
            \includegraphics[width=6.5cm]{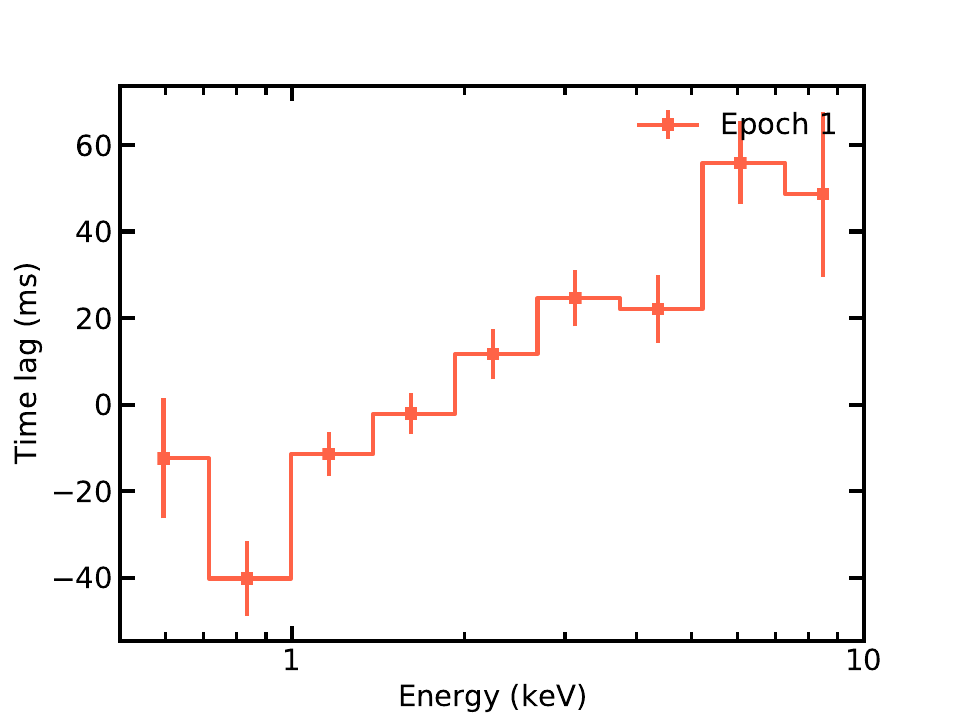}
        \includegraphics[width=6.5cm]{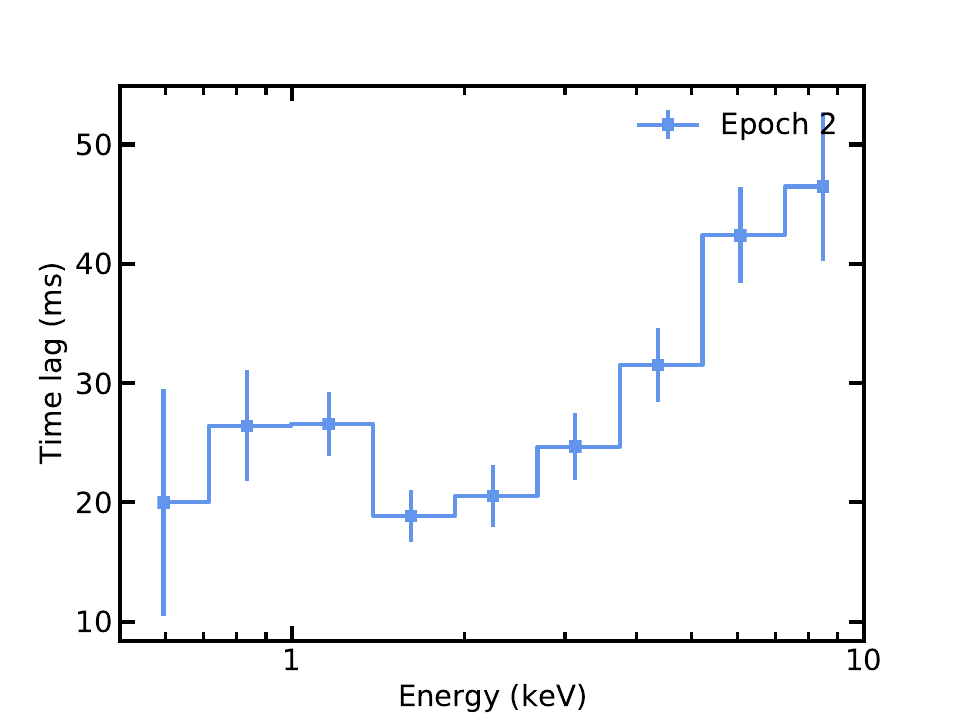}
        \includegraphics[width=6.5cm]{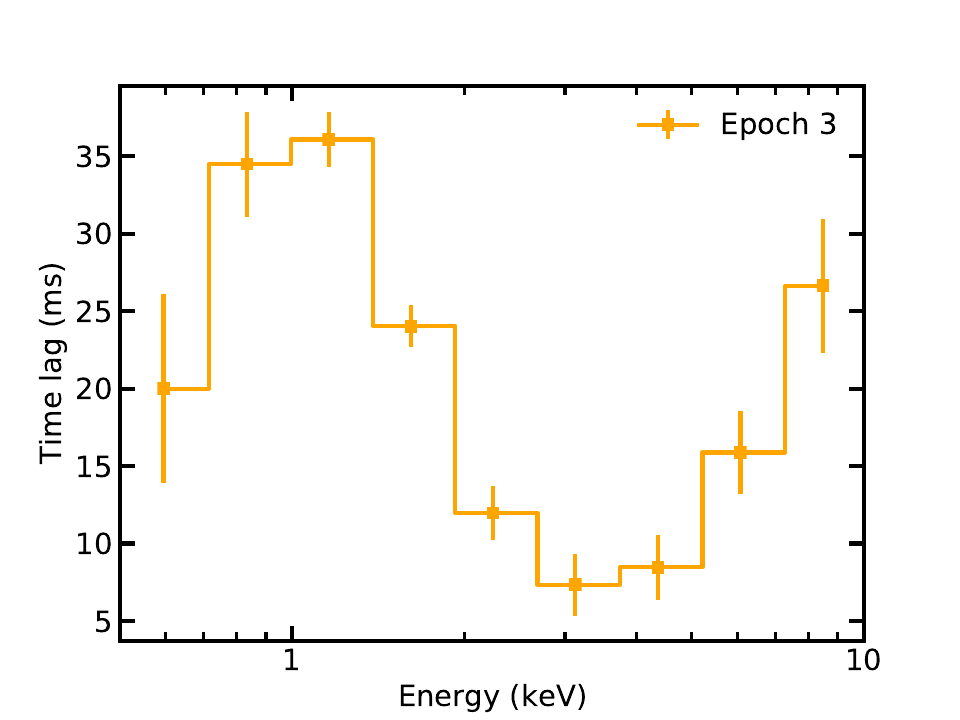}
        \includegraphics[width=6.5cm]{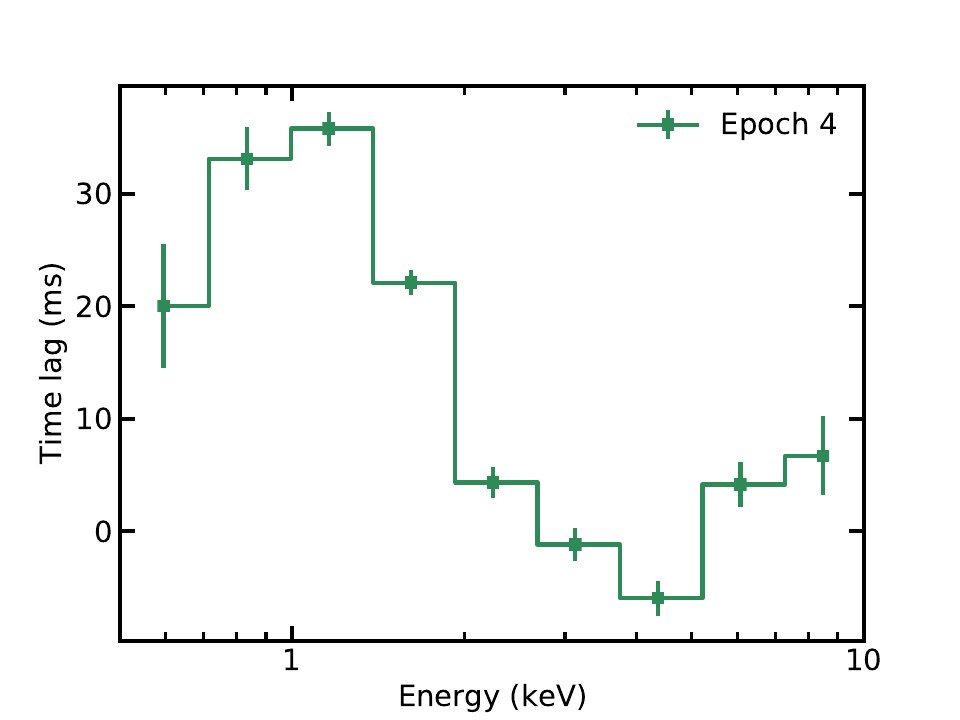}
        \includegraphics[width=6.5cm]{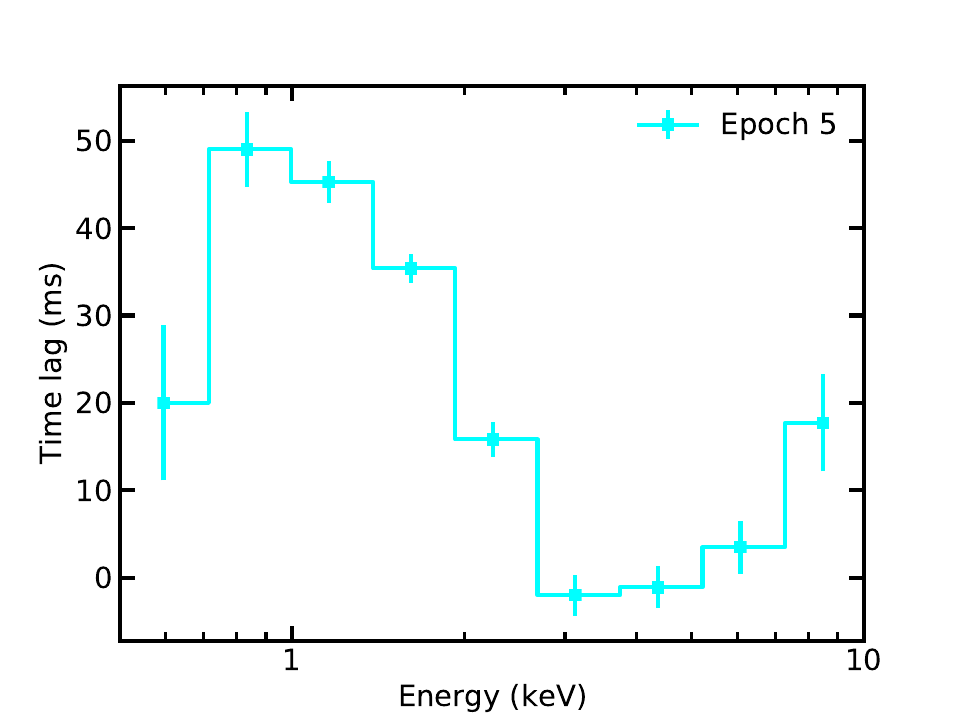}
        \includegraphics[width=6.5cm]{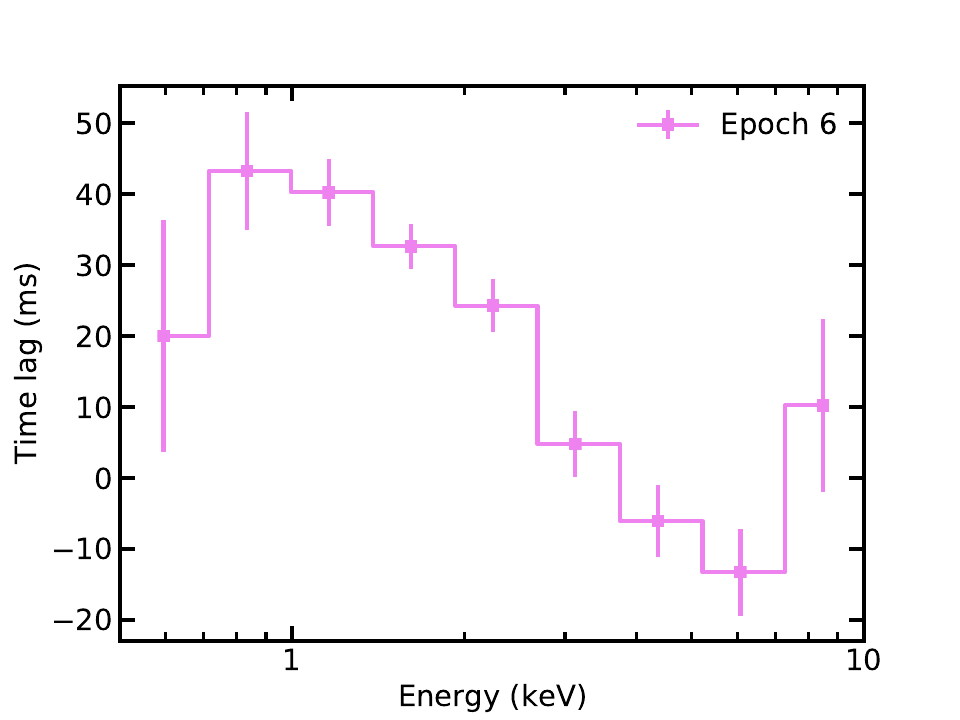}
    \caption{Lag--energy spectra in the 0.1--1 Hz frequency range for the six representative epochs, computed using the same reference band as in Fig.~\ref{fig6}. The spectra evolve from being dominated by hard lags in the LHS to soft-lag-dominated profiles in the HIMS.}

    \label{fig7}
\end{figure*}

At high frequencies (1--10 Hz), the lag--energy spectra display a similar shape at all epochs, characterized by soft lags at low energies that gradually decrease with energy, and hard lags at higher energies. Although no prominent Fe K lag is detected, the soft-energy response and the valley-like lag--energy shape are consistent with thermal reverberation from the disk. The absolute amplitude of the reverberation lags increases from epoch 1 to epoch 2 and then remains approximately constant in the subsequent epochs, following the same evolutionary trend as that observed in the lag--frequency spectra (Fig.~\ref{fig3}).

At low frequencies (0.1--1 Hz), the lag--energy spectra exhibit a markedly different evolution. In the LHS (epochs 1 and 2), the spectra are dominated by hard lags, with the lag amplitude increasing with energy. As the source enters the HIMS (from epoch 3 onward), the lag--energy spectra change significantly and become dominated by soft lags, resembling the high-frequency cases. This transition suggests that the low-frequency hard lags become progressively weaker as the source evolves, allowing the reverberation lags to dominate.

Nevertheless, the low- and high-frequency lag--energy spectra in the HIMS are not entirely identical. In the low-frequency spectra, the lag first increases with energy below $\sim$1 keV and then decreases at higher energies, with $\sim$1 keV marking the turning point. Notably, this feature is not exclusive to the HIMS, but already emerges in the late LHS (epoch 2), where a small hump-like structure around $\sim$1 keV is observed. This feature is likely related to the disk-leading effect, in which long-timescale variations in the disk emission lead those of the Comptonized component. This behavior has been observed in several BHXBs and might reflect the response time of the corona to variability in disk seed photons \citep{uttley2011causal,de2015tracing}. Since the disk-leading component produces a trend opposite to that of reverberation lags in the lag--energy spectrum, their superposition can naturally account for the observed structure at low energies. A more quantitative investigation of this effect would require detailed modeling of the lag--energy spectra, which is beyond the scope of this work.

\section{Discussion}\label{sec4}

\subsection{Observational overview}

From the analysis of \nicer{} observations, the lag behavior of Swift J1727.8--1613 reveals a coherent pattern in the frequency and energy domains. From the lag--frequency spectra, the high-frequency soft lag amplitude increases rapidly during the LHS and then remains approximately constant throughout the HIMS. The low-frequency lags undergo a clear transition from hard lags in the LHS to soft lags in the HIMS, with soft lag amplitudes comparable to those measured at higher frequencies.

These properties are further supported by the lag--energy spectra. At high frequencies, the overall shape of the lag--energy spectra remains stable across all epochs, displaying the characteristic structure expected from disk reverberation. At low frequencies, the lag--energy spectra evolve from being dominated by hard lags to a regime where soft lags become increasingly prominent, similar to the lag-frequency spectra.

\subsection{Origin of the lag evolution}

During the early LHS (epoch 1), the high-frequency soft-lag amplitude is significantly smaller than in later epochs, while the low-frequency hard lags are pronounced. Previous studies have shown that propagation-induced hard lags follow a power-law dependence on Fourier frequency, with larger amplitudes at lower frequencies \citep{nowak1999rossi,pottschmidt2000temporal}. This suggests that when hard lags are strong, they can extend into the frequency range in which soft lags are measured. As a result, the overlap between hard and soft lags can lead to contamination, causing the measured soft-lag amplitude in epoch 1 to be biased toward lower values. This interpretation is further supported by the observed anticorrelation between the amplitudes of the hard and soft lags during the LHS. In this scenario, the intrinsic reverberation lag is partially diluted by the presence of hard lags.

As the source evolves toward the HIMS, the hard lag decreases rapidly, reducing its effect on the high-frequency soft lags. As a result, the soft-lag amplitude increases and quickly reaches a stable value of $\sim$10 ms. As the hard lag continues to decrease, the low-frequency lag is also increasingly dominated by soft lags, with a comparable amplitude. When the source enters the HIMS, the soft-lag amplitude remains stable in all observations and as a function of Fourier frequency. This behavior is consistent with expectations for reverberation delays, which are governed by light-travel times and are therefore largely independent of variability timescales. The fact that comparable soft-lag amplitudes are measured at low and high frequencies further supports the interpretation that the observed lags are dominated by reverberation when the effect of hard lags becomes negligible.

Taken together, these results provide a natural explanation for the apparent increase in the soft-lag amplitude from the early LHS to the HIMS. Instead of reflecting a substantial change in the intrinsic reverberation timescale, the increase can be understood as the gradual reduction of hard-lag contamination, allowing the underlying reverberation lag to emerge more clearly. Although we cannot fully exclude the possibility that the intrinsic reverberation lag evolves to some extent from the LHS to the HIMS, disentangling this evolution from the effect of hard lags remains difficult. A quantitative decomposition of the propagation and reverberation contributions would therefore be valuable, but is beyond the scope of the present work and will require more detailed modeling in future studies.

\subsection{Implications for the accretion geometry and comparison with previous studies}

X-ray reverberation lags are widely used as a probe of the inner accretion geometry in BHXBs because they primarily reflect the light-travel time between the corona and the disk. Variations in the lag amplitude are therefore commonly interpreted as changes in the spatial scale of the accretion flow, such as the evolution of the inner disk radius or the coronal extent.

For Swift J1727.8--1613, the nearly unchanged soft-lag amplitude throughout the HIMS suggests that the average light-travel distance between the corona and the disk remains approximately constant during this state, consistent with a relatively stable disk--corona geometry. This interpretation is also supported by spectral timing analyses of \citet{stiele2024nicer}, which indicate a relatively stable inner disk radius during this phase, as inferred from the \texttt{diskbb} normalization and the evolution of the QPO frequency.

In the LHS, the increase in the soft-lag amplitude can in principle be associated with an evolution of the accretion geometry, such as an expansion of the corona or an increase in the inner disk radius. However, as discussed above, the increase in the measured soft-lag amplitude does not necessarily reflect an intrinsic change in the reverberation timescale, but can instead be attributed to the varying effect of hard lags. Although it is difficult to determine whether the intrinsic soft-lag amplitude (after accounting for this contamination) remains unchanged from the LHS to the HIMS, the results leave the possibility open that the accretion geometry is already relatively stable during the LHS.

The reverberation lag behavior observed in Swift J1727.8--1613 is relatively uncommon among black hole X-ray binaries. By comparing this with the population study of reverberation lags presented by \citet{wang2022nicer}, \citet{ingram2024tracking} noted that this source exhibits a significantly larger lag amplitude in the hard state than most systems and suggested that it might belong to a subpopulation that was underrepresented in previous samples. Our results indicate that in earlier stages of the LHS, the soft-lag amplitude in Swift J1727.8--1613 is comparable to the typical values observed in other black hole X-ray binaries at the level of $\sim$2 ms \citep{wang2022nicer}. However, the reverberation lag measured in the early LHS in Swift J1727.8--1613 is likely affected by contamination from strong hard lags, which can bias the observed amplitude toward lower values. This result motivates a broader consideration of whether hard-lag contamination can also affect reverberation measurements in other black hole X-ray binaries, although dedicated analyses of individual sources are required to evaluate its importance. Several sources, for example, MAXI J1820+070 and GX 339--4, exhibit strong low-frequency hard lags during the hard state without showing a rapid decrease comparable to that observed in Swift J1727.8--1613 \citep{de2015tracing, kara2019corona}. This suggests that the effect of hard-lag contamination on reverberation measurements is more widespread than previously recognized and should be carefully considered when interpreting lag evolution across spectral states.

\subsection{Disappearance of the hard lags}

As shown in Fig.~\ref{fig5}, the amplitude of the low-frequency hard lags decreases rapidly from $\sim$100 ms in epoch 1 to values consistent with zero by epoch 2, after which the low-frequency regime becomes fully dominated by soft lags. Correspondingly, the lag--energy spectra also exhibit a dramatic evolution from the LHS to the HIMS (Fig.~\ref{fig7}).

Hard lags in black hole X-ray binaries are commonly interpreted within the framework of inward-propagating fluctuations in the mass-accretion rate \citep{lyubarskii1997flicker,kotov2001x,arevalo2006investigating, he2025daily}. In this picture, perturbations are generated at large radii and propagate inward through the hot accretion flow, modulating the X-ray emission at different energies. The observed time delays arise from the radial stratification of the emitting regions, often parameterized by an energy-dependent emissivity profile of the form $ \propto r^{-\gamma(E)} $. If higher-energy photons originate preferentially from smaller radii (i.e., $\gamma(E_{\rm h}) > \gamma(E_{\rm s})$), fluctuations will reach the hard-emitting region later, naturally producing hard lags \citep{ingram2012modelling}. Within this framework, the observed decrease in the hard-lag amplitude can be attributed to two effects. One possibility is a reduction in the spatial extent of the corona, which would shorten the propagation path and thus reduce the lag. Alternatively, the difference between the emissivity profiles of the soft and hard bands might decrease, that is, $\gamma(E_{\rm h}) \simeq \gamma(E_{\rm s})$, implying that photons in different energy bands are emitted over similar radial regions and thereby suppress propagation-induced lags.

A comparison between the evolution of low- and high-frequency lag amplitudes (Figs.~\ref{fig4} and \ref{fig5}) provides an important constraint. The reverberation lag reaches a stable value of $\sim$10 ms before the source enters the HIMS, while the low-frequency hard lag still decreases. This suggests that the overall accretion geometry has already reached a relatively stable configuration at an earlier stage, making it unlikely that the observed evolution of hard lags is primarily driven by large-scale geometric changes. Instead, the results favor a scenario in which the radial emissivity profile evolves during the outburst. In particular, a decreasing contrast between the emissivity profiles at different energies would naturally lead to a reduction of propagation-induced lags. This can reflect changes in the energy dissipation within the corona, such as a redistribution of heating or a more homogeneous emission structure. Detailed modeling is required in future work to quantify these effects.

\section{Conclusions}\label{sec5}

We have performed a comprehensive analysis of X-ray time lags in Swift J1727.8--1613 during its 2023 outburst using \nicer{} observations, focusing on their evolution across spectral states.

The high-frequency soft lag increases rapidly from the LHS to the HIMS and then remains approximately constant at $\sim$10 ms throughout the HIMS. In the frequency range in which the soft lag dominates, the lag shows little dependence on Fourier frequency and exhibits a stable lag--energy spectrum characteristic of disk reverberation. On the other hand, the low-frequency lag evolves from hard to soft as the hard lags rapidly disappear, allowing the reverberation signal to be observed over a broad frequency range.

The nearly unchanged reverberation lag during the HIMS suggests that the characteristic light-travel distance between the corona and the disk varies little during this phase, consistent with a relatively stable inner accretion geometry. The apparent evolution of reverberation lag amplitude from the LHS to the HIMS can be largely explained by the diminishing effect of hard lags and is not entirely due to changes in the light-travel distance.

Swift J1727.8--1613 provides a case in which the reverberation signal can be more clearly isolated over a broad frequency range when the hard lags become weak. This offers a more direct view of the underlying reverberation timescale with reduced contamination, providing a complementary perspective on the evolution of accretion geometry in black hole X-ray binaries.

\begin{acknowledgements}

This work made use of data from the \nicer{} mission and the HEASARC online data archive, operated by NASA's Goddard Space Flight Center.
W.Y. acknowledges support from the Alexander von Humboldt Foundation. S.A. acknowledges support from the Bundesministerium für Wirtschaft und Energie through the Deutsches Zentrum für Luft- und Raumfahrt e.V. (DLR) under grant No.~50 OR 2517. Z.-X.Y. acknowledges support from the National Natural Science Foundation of China (NSFC) under Grant No. 12403053 and the Shandong Provincial Natural Science Foundation under Project No. ZR2024QA076. X.F. acknowledges support from the NSFC under Grants Nos. 12322307, 12273026, and 12361131579, and from the “Fundamental Research Funds for the Central Universities”.

\end{acknowledgements}

\bibliographystyle{aa} 
\bibliography{aa} 
\end{document}